\documentclass[a4paper]{jpconf}
\usepackage{graphicx,amssymb}
\begin{document}
\title{Quantum kagome antiferromagnet~: ZnCu$_3$(OH)$_6$Cl$_2$}

\author{Philippe Mendels, Fabrice Bert}

\address{Laboratoire de Physique des solides, Universit\'e Paris-Sud, UMR CNRS 8502, 91405 Orsay Cedex, France}

\ead{philippe.mendels@u-psud.fr}

\begin{abstract}
Herbertsmithite, ZnCu$_3$(OH)$_6$Cl$_2$, is the prototype candidate for a spin liquid behavior on a geometrically perfect kagome lattice. Its discovery and the absence of any evidence for spin-freezing down to the lowest probed temperature to-date enable one to explore the properties of kagome-related physics in an  unprecedented temperature range. We review its properties and discuss some open issues. A tentative comparison to models is also performed.
\end{abstract}

\section{Introduction}\smallskip
%After the original paper by Anderson and Fazekas in 1973-74~\cite{Fazekas} and the revival of their seminal mainstream proposal of a resonating %valence bond ground state in the context of High Temperature Superconductors~\cite{Anderson}, kagome antiferromagnets have been recognized since the %early 90's as the corner stone of the search for a quantum spin liquid state. Strong indications for very original physics had been found in some %experimental systems with a kagome based geometry, such as magnetic freezing occuring only at very low temperatures on the scale of the exchange %energy and fluctuating ground states, two features recognized as the landmarks of frustration~\cite{Ramirez90,Uemura}.

Herbertsmithite ZnCu$_{3}$(OH)$_6$Cl$_2$, a rare mineral recently extracted and named after G.F. Herbert Smith (1872-1953) who discovered the mother atacamite family in 1906 ~\cite{Braithwaite}, has been synthesized for the first time in 2005, in M.I.T.~\cite{Shores}. It has triggered a major renewal in the search of a quantum spin liquid on the kagome Heisenberg antiferromagnetic (KHA) lattice and opened the possibility to directly expose theories to experiments for the KHA ground state which has eluded any definitive conclusion for the last twenty years. Altogether with triangular compounds $\kappa$(ET)$_2$Cu$_2$(CN)$_3$~\cite{Kanoda} and more recently, (ET)Me$_3$Sb[Pd(dmit)$_2$]$_2$~\cite{Itou}, it combines low dimensionality, a quantum character associated with spins 1/2 and the absence of magnetic freezing which was coined as the "end to the drought of spin liquids"~\cite{PLee,Mendels2010}. Herbertsmithite is the first experimental example of a kagome lattice where no order has been found at any temperature well below $J$ through all experimental techniques~\cite{Mendels2007, Helton}. This illustrates the power of the combination of an enhancement of quantum fluctuations for $S=1/2$ spins with the frustration of antiferromagnetic interactions on the loosely connected kagome lattice to stabilize novel ground states of magnetic matter.

Most of the models point to an exotic ground state and two broad classes of spin liquids have been explored theoretically, in particular for the kagome antiferromagnet~\cite{bookHFM}, (i) topological spin liquids which feature a spin gap, a topological order and fractional excitations~\cite{theory1}; (ii) algebraic spin liquids with gapless excitations in all spin sectors~\cite{theory2}.

Here, we partially review some of the most relevant experimental properties of Herbertsmithite, and complement some aspects of our recent review on this compound~\cite{Mendels2010}. We tentatively discuss which model could be consistent with these findings within the context of well identified deviations to the ideal Heisenberg kagome antiferromagnet.
\section{Structure and magnetic couplings}\smallskip

\begin{figure}[h]
\begin{minipage}{8cm}
\begin{center}
\includegraphics[width=6 cm]{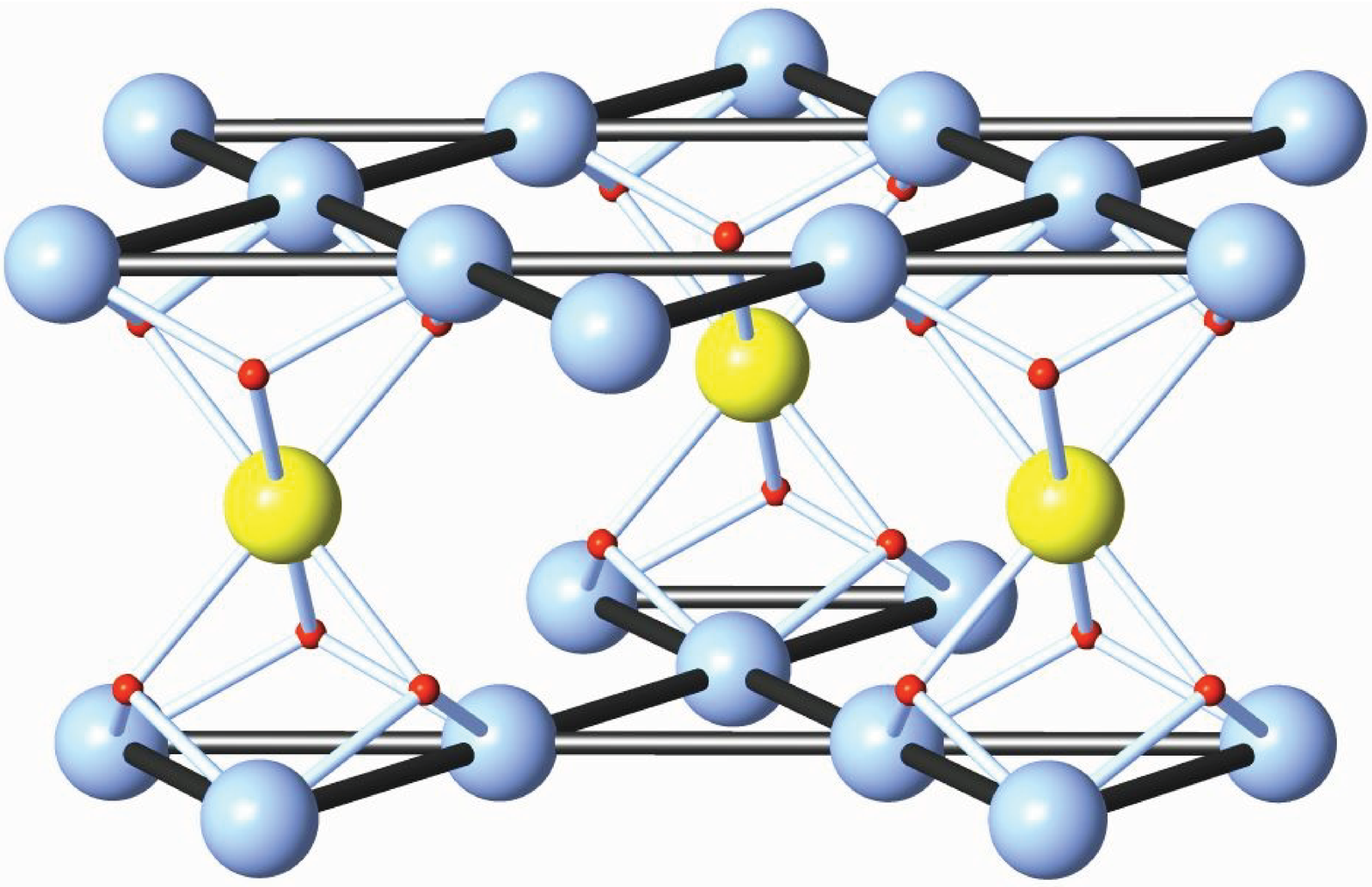}
\caption{\label{fig1a}Simplified structure of ideal Herbertsmithite where only Cu (kagome top and bottom layers), Zn (intermediate triangular layer) and O which bridge between two Cu in the kagome planes, have been represented.}
\end{center}
\end{minipage}\hspace{2pc}%
\begin{minipage}{7cm}
\begin{center}
\includegraphics[width=3.5cm]{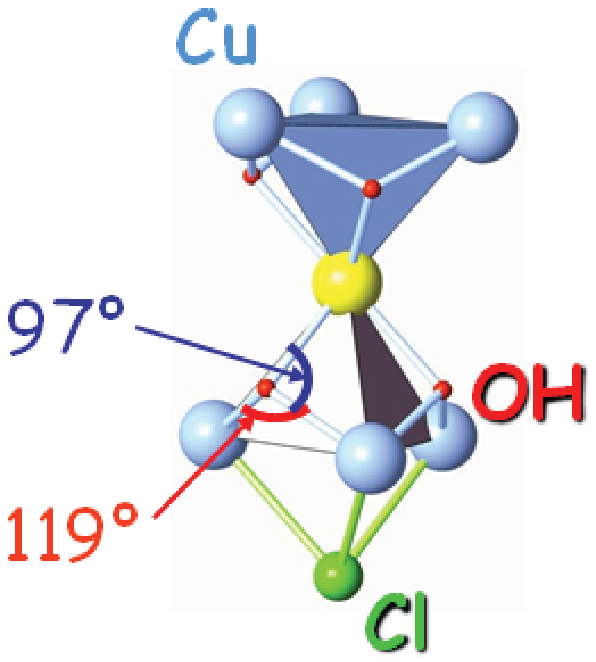}
\caption{\label{fig1b}Local structure emphasizing the coupling path within a kagome layer and between one Cu kagome site and one Cu substituted on the Zn site, named interplane site in the text.}
\end{center}
\end{minipage}
\end{figure}
Herbertsmithite ZnCu$_3$(OH)$_6$Cl$_2$ is the $x=1$ end compound of the Zn-paratacamite family Zn$_x$Cu$_{4-x}$(OH)$_6$Cl$_2$~\cite{Braithwaite,Shores}. It can be viewed as a double variant of the parent clinoatacamite compound ($x=0$), situated at the other end of the family. Starting from the latter, a Jahn-Teller distorted $S=1/2$ pyrochlore, the symmetry first relaxes from monoclinic ($P2_1/n$) to rhombohedral ($R\overline{3}m$) around $x=0.33$, leading to a perfect kagome lattice in the $a$-$b$ plane with \emph{ isotropic} planar interactions; then, in the $c$- elongated $x>0.33$ pyrochlore structure, the magnetic bridge along the $c$-axis between $a$-$b$ kagome planes is progressively suppressed by replacing the apical Cu$^{2+}$ by a diamagnetic Zn$^{2+}$.

%Due to a more favorable electrostatic environment, Cu$^{2+}$ is expected to preferentially occupy the distorted octahedral kagome sites only. When $x=1$ the $S=1/2$  ions should therefore form structurally perfect kagome layers that are themselves well separated by diamagnetic Zn$^{2+}$(Fig.~\ref{structure_Herbert}).

In Herbertsmithite, the $119 ^{\circ}$ Cu-OH-Cu bond-angle yields a moderate planar super-exchange interaction $J\simeq 180(10)$~K in comparison with other cuprates such as High $T_c$'s where the Cu-O-Cu bonding is linear ($J\sim 1000-2000$~K). Note that the former value is consistent with that observed in cubanes where a similar OH bridge links adjacent coppers~\cite{clinoatacamite_Wills2}. %The exchange constant is determined from the fit of the high temperature dc-susceptibility to series expansion calculations for the kagome case~\cite{RigolSingh1, MisguichHerbert}.
Both neutron low-$T$ data on the $x=0$ compound and a simple Curie-Weiss analysis of the variation of the high-$T$ susceptibility with $x$, suggest that the interaction between apical and planar Cu is weakly ferromagnetic and estimated to be $J^{'} \sim 0.1 J$. This is also in line with Goodenough-Kanamori rules which predict a weak ferromagnetic exchange for a Cu-O-Cu angle of $96.7 ^{\circ}$ and with the presence of a weak ferromagnetic component in the frozen phase of Zn depleted ($x < 0.66$) compounds.

\section{Categorizing non-ideality and the related impact on physical properties}\smallskip
The dominant term in the Hamiltonian is certainly the Heisenberg exchange interaction between neighboring spins within the kagome layer (see above). Yet, frustrated antiferromagnets are known to yield non-spin liquid ground states under small perturbations. In the case of Herbertsmithite where both disorder and Dzyaloshinkii-Moriya (DM) anisotropy are at play, the absence of any kind of order or freezing seems to be quite surprisingly robust.
\subsection{Dzyaloshinkii-Moriya interactions}\smallskip
In the absence of any macroscopic high-T susceptibility results published on single crystals, ESR proves to be the most appropriate technique to investigate the local spin anisotropy. The inclusion in the Hamiltonian of a Dzyaloshinkii-Moriya interaction $\overrightarrow{D} _{ij}.\overrightarrow{S}_i \times \overrightarrow{S}_j$ is necessary since there is no inversion symmetry  between two adjacent Cu. $\overrightarrow{D}$ has two components, one perpendicular to the planes, $D_z$, and one planar $D_p$ which act differently on the susceptibility and the specific heat. The effects are quite non linear and the large value of $D_p$ invoked quite early after the discovery of Herbertsmithite to explain the rapid increase of the susceptibility at low $T$~\cite{RigolSingh1,RigolSingh2} is far beyond the limit set by the ESR results. The small measured value of $D_p \sim 0.01 J$ and sizeable $D_z\sim 0.04 - 0.08 J$~\cite{Zorko2008,ElShawish} yield a fairly small correction, even negative, to the susceptibility as compared to the Curie-like upturn. Given this, the simple difference with series expansion at temperatures as high as $J/2$ argues for an extra term dominated by -if not only due to- out of plane defects~\cite{MisguichHerbert}. Definitely, the latter is responsible for the drastic low-$T$ upturn of the susceptibility.

Studies of the phase diagram of a quantum Heisenberg kagome lattice with DM interactions show that an original spin liquid phase can survive, at variance with the classical case where any minute amount of DM anisotropy will lead to long-range order~\cite{Elhajal}. For spins $S=1/2$ a quantum critical point indeed appears for $D/J \sim 0.1$ which separates the liquid and the N\'eel states~\cite{Cepas,Messio,Huh}. A peculiar variation of the susceptibility $\chi \sim T^{-0.53}$ at the QCP is expected~\cite{Huh}.

\begin{figure}[h]
\includegraphics[width=8 cm]{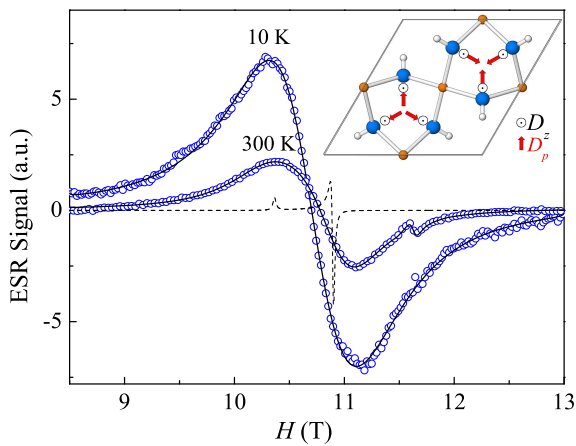}\hspace{2pc}%
\begin{minipage}[b]{16pc}\caption{\label{label}ESR spectra taken at 300 and 10~K. The broad linewidth is due to a large anisotropy. The dashed line shows the anisotropy of the $g$- shift tensor, hence the minute deviation from a pure Heisenberg character of the spins, quite common for Cu$^{2+}$ ions. In the inset the Dzyaloshinskii vectors are represented. The Cu and O are respectively represented by small and large full circles (red and blue online). Adapted from ref.~\cite{Zorko2008}.}
\end{minipage}
\end{figure}

%\begin{figure}[h]
%\begin{center}
%\includegraphics[width=8 cm]{fig3.eps}
%\caption{\label{fig3}ESR spectra taken at 300 and 10~K. The broad linewidth is due to a large anisotropy. The dashed line shows the anisotropy of the $g$- shift tensor, hence the minute deviation %from a pure Heisenberg character of the spins, quite common for Cu$^{2+}$ ions. In the inset the Dzyaloshinskii vectors are represented. The Cu and O are respectively represented by small and large %full circles (red and blue online). Adapted from ref.~\cite{Zorko2008}.}
%\end{center}
%\end{figure}
\subsection{Out-of-plane defects}\smallskip
The existence of out of plane magnetic defects due to the substitution of Cu$^{2+}$ $S=1/2$ onto the Zn$^{2+}$ non magnetic triangular layer has been the subject of controversies for a few years but this issue seems now settled. Based on electrostatic Jahn-Teller effect arguments, Cu$^{2+}$ was indeed expected to preferentially sit in the distorted octahedral environment of the kagome planes. For a perfect stoichiometry of three Cu to one Zn, which seems to be achieved according to Inductively Coupled Plasma Atomic Emission Spectroscopy (ICP-AES) characterization, this would lead to a a perfect Cu$^{2+}$ kagome plane and a non-magnetic Zn$^{2+}$ site in between the planes~\cite{Shores}. There is now a consensus that this argument cannot be used as an all and nothing argument.
 \begin{itemize}
 \item Low-$T$-macroscopic measurements can indeed be described consistently within a scenario where the Zn site is occupied by 15-20\% Cu$^{2+}$. The upturn of the susceptibility~\cite{Helton}, the partial saturation of the high field magnetization measured up to 14 T at 1.7~K~\cite{BertHerbert}, the linear field dependence of a Schottky anomaly in the specific heat~\cite{deVries} are strong indications of a Zeeman splitting of quasi-free $S=1/2$ moments.
 \item The low energy magnetic spectral weight in inelastic neutron scattering shifts linearly with field~\cite{Kim}.
%One of the major controversies came out from the impossibility to set a direct completely convincing proof through structural refinements because the cross sections between Zn and Cu  even using neutron diffraction do not differ much. The major progress.
\item There were long-standing evidences that neutron diffraction spectra from various groups were best refined using a 15-20\% occupation of the Zn site by Cu, which was also acknowledged recently by the MIT group. Yet the shallow minimum of the $R$- factor was casting some doubt on this determination. An additional very convincing argument came from the X-ray anomalous scattering data on a single crystal which allow to better quantify the level of Zn and Cu present at each site, namely 15(2)\% of Cu was consistently found on the inter-kagome site~\cite{McQueen}.
\end{itemize}

\begin{figure}[h]
\begin{center}
\includegraphics[width=15 cm]{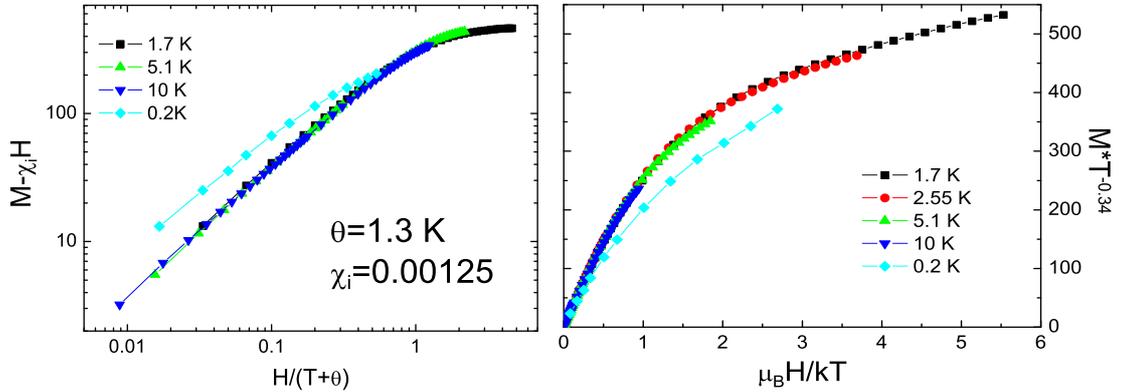}
\caption{\label{fig4}(Color on line) Two scalings for the low-$T$ magnetization, $M$,  measured up to 14 Teslas. The 0.2~K data deviate from the scaling. Left: the contribution from the kagome planes which remains linear at high field has been subtracted. The remaining part is associated with quasi-free spins and the Curie-Weiss like scaling which works well down to $\sim 1$~K, suggests some magnetic interaction between the spins of the order of 1~K. Right:following the idea of a quantum critical behavior developed in~\cite{Helton2},an equally good scaling is obtained. Yet, the fitted part does correspond to the now well established contribution from Cu defects on the Zn site which rules out the adaptability of a DM induced quantum critical scenario $M~T^{-0.53}$ \cite{Huh} for these data.}
\end{center}
\end{figure}

Given this, one has to wonder how these quasi-free spins might couple with the kagome planes and how they might impact on the physics. As stated in the previous section, a small ferromagnetic interaction at least one order of magnitude smaller than the in-plane interaction should be anticipated. Surprisingly, susceptibility data and magnetization experiments at low-$T$ do not show  any sign of such a ferromagnetic interaction but rather point at a weak antiferromagnetic effective coupling  of the order of 1~K between these "lone" spins~\cite{BertHerbert}. Since a simple dipolar coupling would yield too weak an interaction of 10~mK for 15\% $S=1/2$ moments, one is led to assume that this effective AF interaction might reveal some indirect coupling through the Cu's of the kagome planes. Using a reciprocity argument, one might infer that the lone spins have little influence on the kagome planes physics down to $\sim 1$~K.

One should as well note in passing that it has been proposed at odds  to interprete this low-T behavior of the susceptibility as an evidence for a quantum critical behavior under the assumption that it could be safely attributed to the kagome plane physics~\cite{Helton2}. Given the arguments of the previous section - the low-$T$ macroscopic susceptibility is dominated by the inter-plane channel defect-, the latter condition certainly does not apply here. Fig.~4 shows the comparison between a Curie-Weiss analysis of the susceptibility and the scaling proposed in ref.~\cite{Helton2}. They equally work well for $T>1$~K but fail to interprete the 0.2~K data which signals some different physics of the defects in the low-$T$ range. The muon relaxation rate also displays some peculiar change around 1K.  Such an analysis in terms of quantum criticality therefore can be safely replaced by the more sound scenario where a weak effective interaction couples these moments.

\subsection{In-plane vacancies~: an open issue?}\smallskip
While 15-20\% of Cu occupation of the inter-plane site was found, depending on the technique used and likely on the samples, ICP-AES analyses have always reliably supported the idea of a perfect global stoichiometry of 1.00:3.00 Zn/Cu ratio. This leads to the straightforward conclusion that intersite mixing occurs and a 5-7\% amount of Zn$^{2+}$ occupies the kagome planes, then creating spin vacancies.

This seems to be quite disputed by the recent X-ray anomalous scattering data~\cite{McQueen}, where an upper limit of 1(3)\%Zn on the kagome layer was found. On the other hand, pushing to the extreme limit of 5\% error on the 1.00:3.00 Zn/Cu ratio from ICP-AES reported on the same sample, would yield around 4\% Zn on the kagome sites if 15\% of the Zn sites are occupied with copper. We see that we are at the limit of the consistency between the two sets of data taken on the same sample, which likely require further inspections.

\begin{figure}[h]
\begin{center}
\includegraphics[width=10 cm]{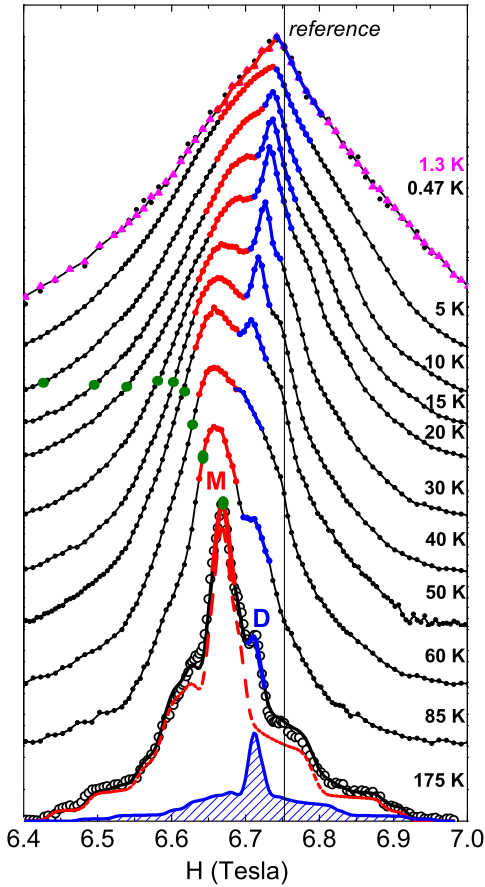}
\begin{minipage}[b]{12pc}\caption{\label{spectraO17}(Color online) $^{17}$O NMR spectra from 175~K to 0.47~K from ref.~\cite{Olariu}. The shift reference is indicated by the vertical line. (M) and (D) refer to the two oxygen sites in the kagome planes as described in the text. Large full green circles indicate the position of the center of the line that would be expected from $\chi_{macro}$. At 175~K, a detailed analysis was performed and two contributions are evident from the displayed fits (see text and section 4). The hatched area represents the contribution from (D) oxygens sitting nearby one Zn defect in the kagome plane.}
\end{minipage}
\end{center}
\end{figure}

Two other arguments based on $^{17}$O NMR spectra give evidence for the existence of in-plane defects~\cite{Olariu}.
\begin{itemize}
\item{Typical spectra are displayed in Fig.~\ref{spectraO17}. From a detailed analysis of the marked singularities of the powder lineshape which positions are governed by the magnetic shift and quadrupolar effects, {\it two different sites} can be identified. The most intense line at high $T$, named main(M), is associated with O sites linked to two Cu sites in the kagome planes. The other line (D) which is very sharp at low-$T$ can be consistently interpreted as coming from O sites linked with one Cu and one non-magnetic Zn in the kagome planes. It is indeed half shifted at high-$T$ as compared to the (M) line and corresponds to 20\% spectral weight, {\it ie} a 5\% Zn occupation of the kagome planes. Its sharpness at low $T$  is an additional indication that it corresponds to very specific sites in the kagome planes, the susceptibility of which is not at all related to the Curie-like contribution of the out of plane defects.}
\item{A broadening of the spectrum of the main line (M) is observed at low $T$. Would the kagome plane be free of any defect, one could still argue that:\newline
     (i) The DM anisotropy could be responsible of this broadening on a powder spectrum. In the absence of any calculated field orientation dependence of the susceptibility due to DM anisotropic interaction, we crudely used the calculated susceptibility for an isolated dimer in the presence of DM interaction performed for SrCu$_2$(BO$_3$)$_2$~\cite{Miyahara}. Since O probes two adjacent sites, the non staggered susceptibility should give birth to a powder shape which maximum relative width would be of the order of $\sim g\mu_B H D^2/J^3$. This represents $\sim 10^{-4}$ contribution to the shift, much too small.\newline
     (ii) The $g$-factor anisotropy ~\cite{Zorko2008} could be responsible for some broadening. Its value $<17\%$ is not enough to explain such a huge low-T broadening and that should scale with the shift which is not the case.\newline
     (iii) Out-of plane defects could be responsible for such a width. An asymmetric broadening $\sim 1/T$ would be expected through a direct coupling of these defects to the planes, which is not the case. Such a coupling is also ruled out by the features of the (D) line reviewed above.\smallskip\newline
    One can therefore safely conclude that the susceptibility in the kagome planes is not homogeneous and the $^{17}$O linewidth is a clear signature of the existence of a \emph{staggered response around spinless defects}.}
\end{itemize}
Based on this conclusion the relative 4:1 intensities of the (M) to (D) line argues in favor of 5\% defects in the kagome planes for the investigated sample. These are likely due to 5\% spin vacancies but one cannot of course discard that the (D) line might result from local structural in-plane distortions due to Cu substitutions on the Zn site. Another honest still open possibility is that the samples investigated in NMR and those studied in X-ray are somewhat different.
\section{Evidences for a gapless ground state}\smallskip
Given the dominant contribution of out of plane defects in macroscopic measurements, only site-selective probes can give some insight into the physics of the kagome planes. There are certainly two ideal probes of the kagome physics, Cu and O nuclei. Unfortunately Cu has never been detected in the whole T-range and the detection of a signal which was reported in~\cite{Imai} is certainly encouraging but which Cu, either in-plane or out-of-plane, is detected is still an open question. O therefore appears to be the most secure probe. From a simple comparison to other probes at high-$T$, where the macroscopic susceptibility is dominated by the kagome planes, O is found to be one order of magnitude at least better coupled than Cl~\cite{Imai} and two orders of magnitude better than implanted muons in $\mu$SR experiments~\cite{Ofer}.
\subsection{Kagome susceptibility from $^{17}$O NMR~\cite{Olariu,Bert09}}\smallskip
\begin{figure}[h]
\begin{center}
\includegraphics[width=16 cm]{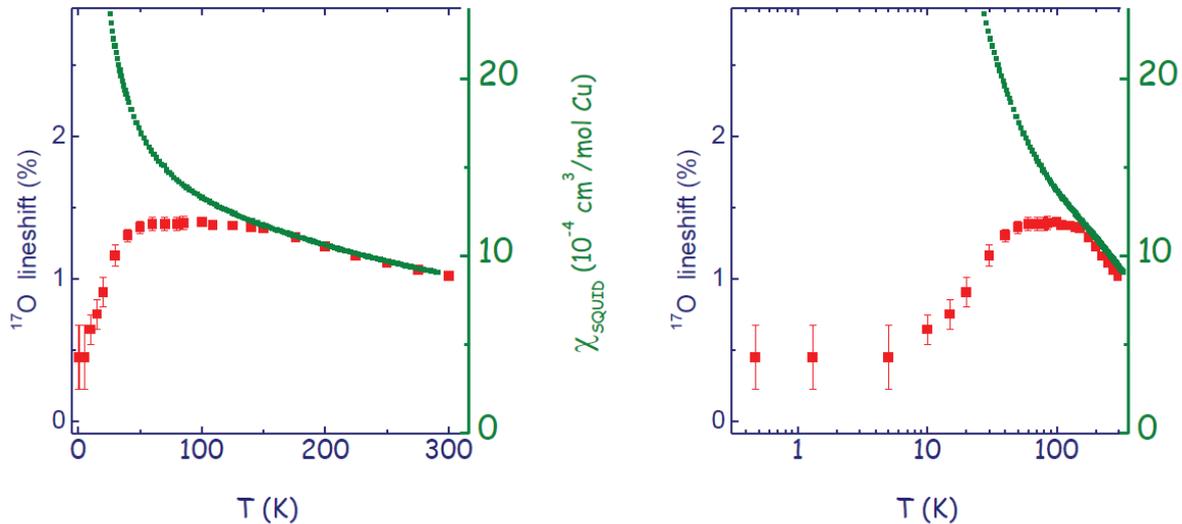}
\caption{\label{fig6}(Color online) Thermal variation of the susceptibility of
the kagome planes measured from the shift of the main $^{17}$O NMR
line (red open symbols) compared to the macroscopic susceptibility
( green solid line). The red dashed line is a guide to the eye for the low-$T$ shift.}
\end{center}
\end{figure}
Tracking the position of the main line gives the shift, hence one can extract the hereafter called "intrinsic" susceptibility  from 300~K$\sim 1.7J$ down to 0.45~K$\sim J/400$. A broad maximum is observed between the high-$T$ Curie-Weiss regime and the low-$T$ regime where the susceptibility monotonously decreases between $J/3$ and $J/20$ to finally level-off at a finite value. The behavior in the intermediate $T$-range can certainly be associated with a strengthening of the short-range antiferromagnetic correlations. Fig.~\ref{fig6} illustrates the huge difference between the intrinsic susceptibility of the kagome planes and the macroscopic susceptibility which is dominated at low $T$ by the inter-plane defects.

Using the DM calculations in a dimer model cited above gives a strong indication that the DM interaction is not in an obvious manner the origin of the non zero susceptibility at $T=0$ and points to a gapless ground state in Herbertsmithite \dots unless for some not obvious reason J should be rescaled to a much weaker effective value, e.g. 0.1 $J$. This calls for further realistic inclusions of the DM interaction into the kagome Heisenberg hamiltonian.

\subsection{Dynamical susceptibility}\smallskip
\begin{figure}[h]
\begin{center}
\includegraphics[width=11 cm]{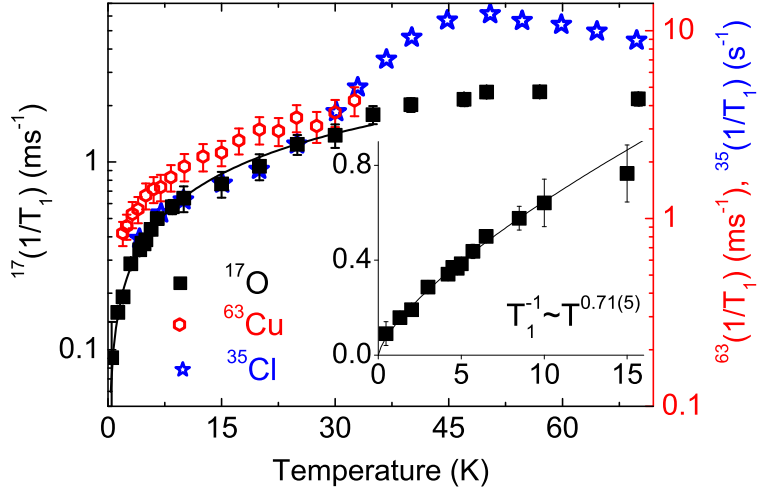}
\caption{\label{fig7}(Color online) Semi-log plot of the relaxation rates obtained for all probe nuclei, $^{17}$O, $^{63}$Cu and $^{35}$Cl. Inset: Linear plot of $T_1^{-1}$ with a power law fit for the low $T$ $^{17}$O data. Adapted from ref.~\cite{Olariu, Imai}.}
\end{center}
\end{figure}

Both NMR $T_1$ relaxation measurements ~\cite{Olariu, Imai} and  inelastic neutron scattering (INS)~\cite{Helton, deVries} give also a strong support to a gapless scenario, at least in the field and temperature ranges which have been probed. NMR $T_1$ measurements taken using various probes agree well from 30~K down to 1.2~K. $^{17}$O data extend down to 0.47 K under an applied field of 6.6~Tesla where a gapless picture is found to hold with a power law behavior $T_1^{-1}\sim T^{0.71(5)}$. Although less sensitive, INS data taken down to 35~mK without any applied field consistently reveal that $\chi''(\omega)$ increases at low $T$ with an uncommon divergent $\omega^{-0.7(3)}$ law.\newline
One should as well note that a very different $T$-behavior is observed in $\mu$SR experiments. A tiny increase of the relaxation rate is found below 1~K, to reach a plateau under zero applied field~\cite{Mendels2007}. Since the dipolar coupling of the muon to nearby moments is an important channel for relaxation, these dynamical effects might be associated with interplane Cu.

\section{A tentative comparison to models}\smallskip
Finding the ground state of the kagome Heisenberg antiferromagnet has represented a "Holy Grail" in the field of frustrated quantum magnetism. Two scenarios have been in balance since the early days, that of a spin liquid, gapped or ungapped, where quantum fluctuations melt away any form of magnetic order and that of a honeycomb valence bond crystal combining dimer bonds within a 36-sites unit cell. The winner is of course that of lowest energy. As an example, the Valence Bond Crystal (VBC) model was found to be the best in 2008 through dimer series expansions~\cite{Singh2007} whereas in late 2010 a Density Matrix Renormalization Group (DMRG) approach strongly supports a spin liquid ground state with a gap, with a possible value of $J/20$~\cite{White}.

On the experimental side, Herbertsmithite does not favor any of these scenarios. Neither a gap is observed much below $J/100$, nor singlets are detected. The central difficulty is to understand better how much the ground state is perturbed by the presence of in-plane defects and DM anisotropy. We have selected a few proposed scenarios and discuss below whether the experimental data might cope with them or not in the realistic experimental situation.
\begin{itemize}
\item \emph{Dirac gapless spin liquid:} Using a fermionic approach in the context of fractional excitations allows to derive a specific heat $C\sim T^2$, a susceptibility $\chi\sim T$, and the NMR relaxation rate $1/T_1 \sim T^\eta$~\cite{Ran2007,Ran2009}. Although the low-$T$ shift roughly scales linearly with $T$ below $J/6$, the unstability of such a spin liquid against DM anisotropy rather argues against such a scenario.
\item \emph{Valence bond crystal/glass:} Nothing in the data collected on Herbertsmithite supports the idea of a simple VBC state~\cite{Singh2007}, which would combine dimer bonds as proposed for the honeycomb 36 sites unit-cell. Including DM interactions certainly relieves the singlet ground state constrain and could lead to some agreement with the experimental data but some effective interaction $\sim 0.1 J$ should be at play to explain the susceptibility, which remains unclear. Interestingly, in presence of disorder, a valence bond glass (VBG) could be stabilized where disorder might restore some susceptibility~\cite{SinghVBG}. In this VBG framework, Raman scattering should give a very characteristic response. The defects are also predicted to yield free spins not simply localized singlets. Yet the recent DMRG calculations seem to disqualify such a valence bond based state, see below.
\item \emph{Gapped spin liquid:} This is the present winner in energy as found in a recent DMRG approach but a gap is predicted of the order of $J/20$ far too high to be consistent with the present experimental findings.
\item \emph{What about the message from exact diagonalizations (ED)?} Besides the increase of the size of samples which can be studied, the computing progress enables now to calculate a large number of levels in any $S$ sector of clusters up to $N=36$ spins. This opens the possibility of a detailed analysis of the magnetism of the cluster. Numerical data of various sizes were recently scrutinized and lead to the conclusion that the singlet-triplet gap  could vanish at the thermodynamic limit although as early as 1998, it was thought to be of the order of $J/20$~\cite{Lecheminant97,Waldtmann98}. The analysis of the minimum energy in each spin sector leads to the conclusion that the KHA could be gapless with a susceptibility consistent with the NMR results, although 4 times larger~\cite{Lhuillier}.

    DM interactions and disorder can also be included in ED calculations~\cite{Dommange03, Rousochatzakis}. As in most models, a singlet is found to localize in the vicinity of a spin vacancy - a relief of frustration on the corresponding triangle - and DM anisotropy again restores a small susceptibility. The longer distance response is staggered. Both features reproduce qualitatively the NMR spectra which respectively signal a specific low-$T$ response in the vicinity of the spin vacancy and a broadening of the line which cannot be ascribed only to DM interactions without defects.

\end{itemize}

\section{Conclusion}
\begin{figure}[h]
\begin{center}
\includegraphics[width=14 cm]{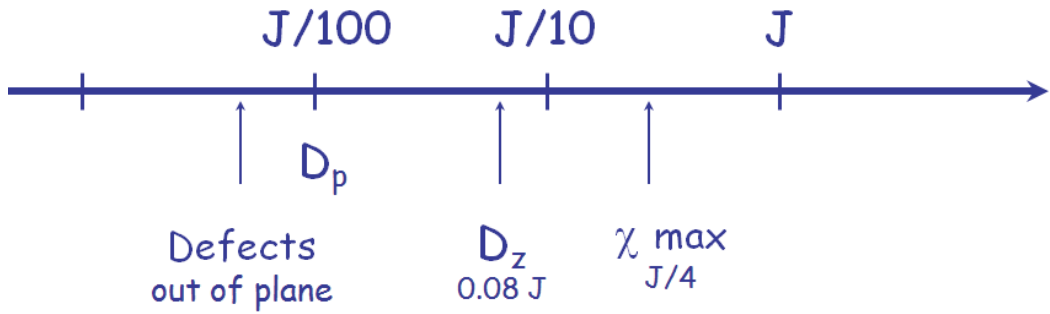}
\caption{Plot of the various energy scales at play in Herbertsmithite.}
\end{center}
\end{figure}
In this paper, we aimed at clarifying some features associated with the defects. Out of plane defects seem not to affect much the physics of the kagome plane but they certainly obscure the macroscopic response. Including DM interaction in models, controlling the in-plane defects in order to extrapolate to a perfect kagome lattice are certainly the way to go for joining theory and experiment. At the present stage, the major result is that Herbertsmithite is a gapless spin liquid under the field and temperature conditions used up to now in experiments. Once the prehistorical age of Herbertsmithite will be over, spinless vacancies might play the role of a smoking gun for the physics of these compounds. Certainly much work is needed to reach that stage and the first synthesis of millimetric crystals are encouraging news for future research in the field \dots and  to the most pessimistic reader who would be tempted to argue that Herbertsmithite is too disordered to shed light on the KHA ground state, EDs give a rather contrasting view that many experimental results can already be reproduced at least at a qualitative level! We certainly are at the stage where the complexity of the compound is well appreciated and one  can now expect some considerable step forward in the materials improvement and the physical discussions on this compound.

\end{document}